\documentclass[12pt,a4paper]{article}
\begin{document}
\sloppy
\title{{\bf Advances in String Theory in \\ Curved Backgrounds : A Synthesis Report}}
\author{Norma.G. SANCHEZ \\
Observatoire de Paris, LERMA\\61, avenue de l'Observatoire\\75014 Paris, 
FRANCE\\Norma.Sanchez@obspm.fr}
\date{ }
\maketitle

\begin{center}
{\bf Abstract :} 
\end{center}
\noindent
A synthetic report of the advances in the study 
of classical and quantum string dynamics in curved backgrounds is provided, 
namely : the new feature of Multistring solutions ; the mass 
spectrum of Strings in Curved backgrounds; The effect of a Cosmological 
Constant and of Spacial Curvature on Classical and Quantum Strings; 
Classical splitting of Fundamental Strings; The General String Evolution in 
constant Curvature Spacetimes; The Conformal Invariance Effects; Strings on 
plane fronted and gravitational shock waves, string falling on spacetime 
singularities and its spectrum. \\
New Developments in String Gravity and String Cosmology are reported : String 
driven cosmology and its Predictions; The primordial gravitational wave 
background ; Non-singular string cosmologies from Exact Conformal Field 
Theories; Quantum Field Theory, String Temperature and the String Phase of 
de Sitter space-time; Hawking Radiation in String Theory and the String 
Phase of Black Holes; New Dual Relation between Quantum Field Theory regime 
and String regime and the ``QFT/String Tango''; New Coherent String States 
and Minimal Uncertainty Principle  in string theory. 
\begin{center}
{\bf CONTENTS}
\end{center}
\noindent
{\bf 1} - Multistring Solutions : a new feature for strings in curved 
backgrounds\\ \\
{\bf 2} - Mass Spectrum of Strings in Curved Spacetimes\\ \\
{\bf 3} - The Effect of a Cosmological Constant and of Spatial Curvature on  
Classical and Quantum Strings\\ \\
{\bf 4} - Classical Splitting of Fundamental Strings\\ \\
{\bf 5} - The General String Evolution in Constant Curvature Spacetimes\\ \\
{\bf 6} - The Conformal Invariance Effects \\ \\
{\bf 7} - Strings on plane fronted and shock gravitational waves, the falling
\\
    on strings on space-time singularities and its spectrum\\ \\
{\bf 8} - Minimal String driven Cosmology and its Predictions\\ \\
{\bf 9} - The primordial Gravitationaal Wave Background in String Cosmology\\ \\
{\bf 10}- Non-Singular String-Cosmologies From Exact Conformal Field Theories\\ \\
{\bf 11}- Quantum Field Theory, String Temperature and the String Phase \\
    of De Sitter Space-Time\\ \\
{\bf 12}- Hawking Radiation in String Theory and the String Phase of Black 
Holes\\ \\
{\bf 13}- New Dual Relation between Quantum Field Theory Regime and String 
    Regime in Curved Backgrounds\\ \\
{\bf 14}- New Coherent String States and Minimal Uncertainty Principle in 
    String Theory\\ \\
In 1987, we started a programme \cite{vs1} to study the string dynamics in 
curved
spacetime and its associated physical phenomena. This study revealed new
insights and new physical phenomena with respect to string propagation in 
flat spacetime (and with respect to quantum fields in curved spacetime). 
The results are relevant both for fundamental (quantum) 
strings and for cosmic strings, which behave essentially in a classical way. 
Approximative and exact solving methods have been developed. Classical and 
quantum string dynamics have been investigated in black hole spacetimes,
cosmological backgrounds, cosmic string spacetime, gravitational wave 
backgrounds, supergravity backgrounds (which are necessary for fermionic
strings), and near spacetime singularities. Physical phenomena like 
classical string instability and non oscillatory motion in time, quantum 
particle transmutation, string scattering, string stretching, have been 
found. For the results (1987-1994) see for example our Chalonge Erice Lectures 1989-1994 and references therein. See (http://www.obspm.fr/chalonge), 
and ``String Theory in Curved Space Times'', Ed. N. Sanchez, WSPC, (1998). \\

\section{Multistring Solutions: A new feature for strings in curved 
backgrounds} 
The discovery of the multistring property \cite{cvms} \cite{vms} in the 
propagation of strings
in curved spacetimes is the consequence of several developments : \\ \\
(i) The classical string equations of motion plus the string constraints were
shown to be exactly integrable in D-dimensional De Sitter spacetime and 
equivalent to a sinh-Gordon model with a Hamiltonian unbounded from below \cite{vs2}.
 Generalization of this result including the Cosh-Gordon and Liouville 
equations, for strings and multistrings in constant curvature spacetimes have 
been given in ref \cite{ls5}.\\ \\
(ii) \begin{em}
Exact 
\end{em}
string solutions were systematically found by soliton methods using the linear system associated to the problem (the so-called dressing method in soliton
theory) \cite{cvms}. In particular, exact circular string solutions were found in 
terms of elementary \cite{vms} and elliptic functions \cite{vls1}.\\ \\
(iii) All these solutions describe one string, several strings or even an
infinite number of different and independent strings. A single world-sheet
simultaneously describes 
\begin{em}
many
\end{em}
different strings. This new feature appears as a consequence of the coupling
of the strings with the spacetime geometry. Here, interaction among the 
strings (like splitting and merging) is neglected, the only interaction is
with the curved background. \\
Different types of behaviour appear in the multistring solutions. For some of 
them the energy and proper size are bounded (``{\it stable strings}'') while 
for
many others the energy and proper size blow up for large radius of the 
universe (``{\it unstable strings}''). \\ \\
In all these works, strings are {\it test} objects propagating on the given
fixed backgrounds. The string energy momentum tensor was computed and the 
string equation of state {\it derived} from the string dynamics in cosmological and black hole spacetimes. Strings obey the perfect fluid relation 
\begin{displaymath}
 p=(\gamma-1)\rho 
\end{displaymath} 
with three different behaviours :\\ (i) {\it Unstable} for 
large R, with negative pressure;\\ (ii) {\it Dual} for small R, with positive
pressure, (as radiation);\\ (iii) {\it Stable} for large R, with vanishing 
pressure, (as cold matter). \\ \\
We find the {\it back reaction effect} of these strings on the spacetime \cite{vs3}. 
This is achieved by considering {\it selfconsistently} the strings as matter 
sources for the Einstein equations, as well as for the complete effective 
string equations, for cosmological spacetimes at the classical level. The
selfconsistent solution of the Einstein-Friedman equations for string 
dominated universes exhibits the realistic matter dominated behaviour for 
large times and the radiation dominated behaviour for early times. That is, 
the {\it standard cosmological} evolution is well generated by strings. It 
must be noticed that there is no satisfactory derivation of inflation in the
context of the effective string equations. De Sitter universe {\it does not} 
emerge 
as solution of the effective string equations. The effective string action 
(whatever be the dilaton, its potential and the central charge term) is not 
the appropriate framework in which to address the question of string driven 
inflation. \\ \\
More recently, \cite{ls1}, new classes of exact {\it \bf multistring} solutions were found. The multistring solutions were classified and their physical 
properties described.\\ In Anti de Sitter spacetime, the solutions describe an 
{\it \bf infinity} number of infinitely long stationary strings of equal energy but different pressures. In De Sitter spacetime, outside the horizon, 
they describe infinitely many {\it \bf dynamical} strings, infalling non radially, scattering at the horizon and going back to spatial infinity in 
different directions. For special values of the constant of motion, there are 
families of solutions with {\it \bf selected finite} numbers of different and independent strings. The strings appear {\it \bf distributed} in 
{\it \bf packets}, the number of strings in each packet, the number of 
``turns'' or ``festoons'' in each string is precisely determined and solely 
dictated by the dynamics, exactly solved in terms of elliptic functions. \\ \\
In Black hole spacetimes, (without cosmological constant) no multistring 
solutions are found \cite{ls2}. In the Schwarzschild black hole, inside the 
horizon, 
the string infalls, with {\it \bf indefinitely} growing size and energy, into the r=0 singularity and the string motion stops there.\\ In the (2+1)- black hole anti-de Sitter background, the string stops at r=0 with {\it \bf  
finite} length \cite{ls3}; the reason being that the point r=0 is not a strong 
curvature singularity in the (2+1)-black hole anti- de Sitter spacetime. 
Outside the horizon, in this spacetime, the multistring solution describes 
infinitely many, infinitely long open strings. 
\section{The String Mass Spectrum in the presence of Cosmological Constant}
The string mass spectrum in the presence of a cosmological constant (for both 
de Sitter and Anti de Sitter spacetimes) was found in ref.\cite{ls3}, 
\cite{vls2}. 
New features as 
compared to the string spectrum in flat spacetime appear, as a {\it \bf 
fine structure effect} (splitting of levels) at all the states beyond the 
graviton, (in both de Sitter (dS) and AntideSitter (AdS) spacetimes), and the 
{\it \bf absence} of a critical Hagedorn temperature in AdS spacetime 
(the partition function for a gas of strings in AdS spacetime is well defined 
for all temperature).\\ \\ The presence of a cosmological constant reduces 
(although do not totally removes) the degeneracy of states as compared with 
flat Minkowski spacetime. In AdS spacetime, the density of states $\rho$(m) 
grows like $\exp$[($\Lambda$ m)$^\frac{1}{2}$], (while, as is known, in 
Minkowski spacetime, $\rho$(m) grows like the Exponential of the mass m). \\ \\
The 
high mass spectrum changes drastically with respect to flat Minkowski spacetime. The level spacing {\it \bf 
grows} with the eigenvalue of the number operator, N, in AdS spacetime, while 
is approximatively constant (although smaller than in Minkowski spacetime and 
slightly decreasing) in dS spacetime. \\ \\ There is an infinite number of 
states 
with arbitrarily high mass in AdS space time, while in dS there is a 
{\it \bf finite} number of oscillating states only.\\The string mass has 
been expressed in terms of 
the Casimir operator C = L$_{\mu\nu}$ L$^{\mu\nu}$ of the O(3,1) De Sitter 
group [O(2,2) group in Anti deSitter \cite{vls2}. See also section 3 below.
\section{Spatial Curvature Effects}
The effects of the spatial curvature on the classical and quantum string 
dynamics is studied in ref.\cite{ls4}. The general solution of the circular 
string motion in static Robertson-Walker spacetimes with closed or open 
sections has been found \cite{ls4}. 
This is given closely and completely in terms of elliptic functions.\\
The {\it \bf back reaction effect} of these strings on the spacetime is 
found : the self-consistent solution to the Einstein equations is a spatially 
closed ({\bf K}$>$0) spacetime with a selected value of the curvature index {\bf K}, {\bf K} = ({\bf G}/${\bf \alpha}$')$^
\frac{2}{3}$ (the scale factor is normalized to unity). No self-consistent solutions with {\bf K}$<$0 exist. \\ \\ We semiclassically quantize the circular 
strings and find the mass m in each case. For {\bf K}$>$0, the very 
massive strings, oscillating on the full hypersphere, have 
\begin{displaymath}
{\bf m}^2 \sim {\bf K} {\bf N^2}, ({\bf N} {\bf \epsilon}  
{\bf N}_0 )
\end{displaymath}
 {\it \bf independent} of ${\bf \alpha}$' and the 
level spacing {\it \bf grows} with n, while the strings oscillating on 
one hemisphere (without crossing the equator) have 
\begin{displaymath}
{\bf m}^2 \sim {\bf \alpha}' {\bf N}
\end{displaymath}
 and a {\bf finite} number of states 
N $\sim$ 1/(K${\bf \alpha}$'). \\ \\For K $<$0, there are infinitely many 
strings states with masses
\begin{displaymath} 
{\bf m}\,{\bf \log}\, {\bf m} \sim {\bf N},
\end{displaymath} 
that is the level spacing grows {\it \bf slower} than {\bf N}.\\The stationary string solutions as well as the generic 
string fluctuations around the center of mass are also found and analyzed in 
closed form.
\section{Classical String Splitting}
We find exact solutions of the string equations of motion and constraints 
describing the classical splitting of a string into two \cite{vmms}. For 
the same Cauchy data, the strings which split have smaller action that the 
string without splitting. This phenomenon is already present in flat space-
time. The splitting process takes place in real (lorentzian signature 
spacetime).\\ \\The solutions in which the string splits are perfectly natural 
within the classical theory of strings. There is no need of extra interactions, (nor extra terms in the action to produce splitting). The difference with the 
non splitting solutions is on the boundary conditions. \\ \\The mass, energy and momentum carried out by the strings are computed. We show that the splitting 
solution describes a natural decay process of one string of mass {\bf M} 
into two strings with a smaller total mass and some kinetic energy. The 
standard non-splitting solution is contained as a particular case. \\ \\We 
also described 
the splitting of a closed string in the background of a singular gravitational 
plane wave, and showed how the presence of the strong gravitational field 
increases (and amplifies by an overall factor) the negative difference between the action of the splitting and non-splitting solutions.
\section{General String Evolution in Constant Curvature Space-Times}
In ref. \cite{ls5}, we have found that the fundamental quadratic form of the 
classical string propagation in (2+1)-dimensional constant curvature 
spacetimes, solves the sinh-Gordon equation, the cosh-Gordon equation, or the 
Liouville equation. In both de Sitter and anti-de Sitter spacetimes, (as well 
as in the 2+1 black hole anti-de Sitter spacetime), {\it all} three equations 
must be included to cover the generic string dynamics. This is particularly 
enlightening since {\it generic} properties of the string evolution can be thus {\it directly} extracted from the properties of these three equations and 
their associated Hamiltonians or potentials, {\it irrespective of any solution}. \\These results complete and generalize our previous results on 
this topic since 
(until now, only the sinh-Gordon sector in de Sitter spacetime was known). \\ 
\\
We also construct new classes of multistring solutions, in terms of elliptic 
functions, to all three equations in both de Sitter and anti de Sitter 
spacetimes, which generalize our previous ones. \\These results can be 
straighforwardly generalized to constant curvature spacetimes of arbitrary 
dimension, by replacing the sinh-Gordon equation, the cosh-Gordon equation, and the Liouville equation by their higher dimensional generalizations. \\ \\
Our results indicate the existence of various kinds of dualities relating the 
different sectors and their solutions in de Sitter and anti-de Sitter 
spacetimes : in the sinh-Gordon sector of de Sitter spacetime, small strings 
are dual (that is, under S $\rightarrow$ 1/S, S being the proper string size, they are mapped to large strings). And, similarly, in the sinh-Gordon sector of 
anti-de Sitter spacetime. Furthermore, in the cosh-Gordon sector, small (large) strings in de Sitter spacetime are dual to large (small) strings in the 
anti-de Sitter spacetime.
\section{Conformal Invariance Effects}
Classical and quantum strings in the conformally invariant background 
corresponding to the $SL(2R)$ WZWN model has been studied in ref \cite{vls3}. 
This background is locally anti-de Sitter spacetime with non-vanishing 
torsion. Conformal invariance is expressed as the torsion being parallelizing; 
and the precise effect of the conformal invariance on the dynamics of both 
circular and generic classical strings has been extracted \cite{vls3}. \\ 
In 
particular, the conformal invariance gives rise to a repulsive interaction of 
the string with the background which precisely cancels the dominant attractive 
term arising from gravity. \\ \\
We perform both semi-classical and canonical string quantization, in order to 
see the effect of the conformal invariance of the background on the string 
mass spectrum. Both approaches yield that the high-mass states are governed 
by 
\begin{displaymath}
m \sim HN (N \in N_0, N ``\mathrm {large}''), 
\end{displaymath}
where {\it m} is the string mass and {\it H} is the Hubble constant. \\
It follows that the level spacing grows proportionally to N :
\begin{displaymath}
\frac {d (m^2 \alpha ^{'})}{dN} \sim N, 
\end{displaymath}
while the string entropy goes like 
\begin{displaymath}
S \sim \sqrt{m}.
\end{displaymath}
Moreover, it follows that there is no Hagedorn temperature, so that the 
partition function is well defined for any positive temperature. \\ \\
All results are compared with the analogue results in anti-de Sitter spacetime, which is a nonconformal invariant background.\\It appears that conformal 
invariance 
{\it simplifies} the mathematics of the problem but the physics remains mainly 
{\it unchanged}. Differences between conformal and non-conformal backgrounds 
only appear in the intermediate region of the string mass spectrum, but these 
differences are minor. For low and high masses, the string mass spectra in 
conformal and non-conformal backgrounds are identical. \\ Interestingly enough, 
conformal invariance fixes the value of the spacetime curvature to be 
$-69/(26\alpha ^{'})$. \\ \\
It has been known for some time that the SL(2,R) WZWN model reduces to 
Liouville theory. In ref \cite{ls6} we give a direct and physical derivation 
of this result based on the classical string equations of motion and the 
proper string size. This allows us to extract precisely the physical effects of the metric and antisymmetric tensor, respectively, on the {\it exact} string 
dynamics in the SL(2,R) background. Also the general solution to the proper 
string size has been found \cite{ls6}. \\ \\
We show that the antisymmetric tensor (corresponding to conformal invariance) 
generally gives rise to repulsion, and it precisely cancels the dominant 
attractive term arising from the metric. Both the sinh-Gordon and the 
cosh-Gordon sectors of the string dynamics in non-conformally invariant AdS 
spacetime reduce here to the Liouville equation (with different signs of the 
potential), while the original Liouville sector reduces to the free wave 
equation. \\
Only the very large classical string size is affected by the torsion. Medium 
and small size string behaviors are unchanged. \\
We also find illustrative classes of string solutions in the SL(2,R) background: dynamical closed as well as stationary open spiralling strings, for which 
the effect of torsion is somewhat like the effect of rotation in the metric. 
Similarly, the string solutions in the 2+1 BH-AdS background with torsion and 
angular momentum are fully analyzed \cite{ls6}.
\section{Strings on plane waves and shock waves. The 
falling of strings  on space-time singularities and its spectrum}
In ref. \cite{vs4}, we studied the dynamics of strings near spacetime 
singularities. \\
We considered plane fronted gravitational-wave backgrounds with a singularity 
of the type $| U | ^{-\beta }$, $U$ being a null coordinate. The case with 
a $\delta (U)$ shock-wave singularity turns out to be similar to the 
$\beta $ = 1 case. \\
New features in the string behavior appear : when $\beta \ge $ 2, the string 
does not propagate through the gravitational wave and it escapes to 
infinity grazing the singularity plane $U$ = 0; one tranverse coordinate 
does not oscillate in time (neither classically nor quantum mechanically) 
and the tunnel effect does not take place. \\ \\
The expectation value of the mass squared $\langle M^2 _> \rangle $ and mode 
number $\langle N_> \rangle $ operators and of the energy-momentum tensor 
are computed. When the transverse size ($\rho _0 $) of the gravitational-
wave front is infinite, divergences in $\langle M^2 _> \rangle $ and 
$\langle N_> \rangle $ appear for $1 \le \beta < 2$ and 3/2 $\le \beta <2$, 
respectively. \\
The short-distance spacetime singularity at $U$ = 0 is not responsible for 
these divergences, but the infinite amount of energy carried by the 
gravitational wave when $\rho _0 = \infty $. \\ \\
In summary, the propagation of strings through these singular spacetimes 
is proven to be physically meaningful for $\beta \ge $ 2 and $\beta <$ 1.  
And this is also the case for 1 $\le \beta < $ 2. \\ \\
In conclusion, test strings do propagate consistently in singular plane 
wave space-times and in shock-wave space-times. We recall that the 
Klein-Gordon equation (for a point particle) is ill defined in this geometry, 
whereas the string equations are well behaved \cite{vs5}, \cite{vs6}. 
Analogous conclusions hold for quantum strings in the Schwarzschild 
geometry where a regular behavior was found at the horizon and at the 
$r$ = 0 singularity \cite{vs7}. That is, strings feel the space-time 
singularities much less than point particles. \\
Furthermore, we would not be surprised by the presence of space-time 
singularities in string theory as long as one sticks to a geometry 
description using a metric tensor $G_{AB}(X)$ (in spite of the fact that 
it fulfills the string-corrected Einstein equations). We do not expect that 
a space-time description in terms of a Riemannian manifold with local 
coordinates $X^A$ will be meaningful at the Planck scale. \\ \\
In ref. \cite{vs8} we fully investigated at the quantum level the nonlinear 
transformation relating the string operators (zero modes and oscillators) 
and Fock space states before and after the collision with gravitational 
shock waves. This throws light on the r$\hat \mathrm{o}$le of the space-time 
geometry in this problem. The treatement was done for a general shock wave 
space-time of any localized source. We computed the {\it exact} 
expectation values of the total number (N) and mass (M$^2$) operators and 
show that they are \underline{finite}.\\ \\ We study the energy-momentum tensor 
of the string and compute the exact expectation values of all its 
components. We analyze vacuum polarization and quadratic fluctuations. All 
these physical magnitudes are {\it finite}.\\ \\ We express all these
quantities in 
terms of exact integral representations in which the role of the real pole 
singularities characteristic of the tree level string spectrum (real mass 
resonances) are clearly exhibited. The presence of such poles is not at all 
related to the structure of the space-time geometry (which may or may not 
be singular). Claims on the divergences of the mass $\langle  M^2 \rangle $ 
and total number $\langle N \rangle $ on these backgrounds(G.T.Horowitz and A.R. Steiff, Phys. Rev. Lett. 64, 260 (1990), Phys. Rev. D42, 1950 (1990)) 
totally overlocked this problem.  
\section{Minimal String Driven Cosmology and its Predictions}
In refs. \cite{is1}, \cite{is2} we constructed a minimal model for the Universe evolution fully extracted from
effective String Theory. \\ \\
By linking this model to a minimal but well established observational 
information, we proved that it gives realistic predictions on early and current energy density and its results are compatible with General Relativity. \\ \\
Interestingly enough, this model predicts the current energy density 
Omega=1 
and a lower limit Omega larger or equal 4/9. On the other hand, 
the energy density at the exit of inflationary stage is also predicted 
$\mathrm{Omega} _{\mathrm{infl}}$ = 1.\\
This result shows  agreement with General Relativity (spatially flat metric 
gives critical energy density) within an unequivalent Non-Eistenian context 
(string low energy effective equations).\\ \\
The order of magnitude of the energy density-dilaton coupled term at the 
beginning of radiation dominated stage agrees with GUT scale.\\ \\
Whithout solving the known problems about higher order corrections and 
graceful exit of inflation, we find this model closer to the observational 
Universe properties than the current available string cosmology scenarii.\\ \\
At a more fundamental level, this model is by its construction close to the 
standard cosmological evolution, and it is driven selfconsistently by the 
evolution of the string equation of state itself. \\ \\
The inflationary String Driven stage is able to reach an enough amount of 
inflation, describing a Big Bang like evolution for the metric.\\
\section{The Primordial Gravitational Wave Background in String Cosmology}
In ref.\cite{is1} we found the spectrum P(w)dw of the gravitational wave background produced in 
the early universe in string theory.\\ \\
We work in the framework of String Driven Cosmology, whose scale factors are computed with the low-energy effective string equations as well as selfconsistent solutions of General Relativity with a gas of strings as source.\\ \\
The scale factor evolution is described by an early string driven inflationary stage with an instantaneous transition to a radiation dominated stage and 
successive matter dominated stage. This is an expanding string cosmology 
always running on positive proper cosmic time.\\ \\
A careful treatment of the scale factor evolution and involved transitions is made. A full prediction of the power spectrum of gravitational waves without 
any free-parameters is given.\\ \\
We study and show explicitly the effect of the dilaton field, characteristic 
to this kind of cosmologies.\\ \\
We compute the spectrum for the same evolution description with three 
differents approachs.\\ \\
Some features of gravitational wave spectra, as peaks and asymptotic 
behaviours, are found direct consequences of the dilaton involved and not 
only of the scale factor evolution.\\ 
\section{Non-Singular String-Cosmologies From Exact Conformal Field Theories}
In ref. \cite{vls} we constructed non-singular two and three dimensional string cosmologies  
using the exact conformal field theories corresponding to SO(2,1)/SO(1,1) and 
SO(2,2)/SO(2,1) coset models.\\ \\
All semi-classical curvature singularities are canceled in the exact theories 
for both of these cosets, but some new curvature singularities 
emerge in the quantum models.\\
However, considering different patches of the global manifolds, allows the 
construction of non-singular spacetimes with cosmological interpretation.\\ \\
In both, two and three dimensions, we constructed non-singular oscillating 
cosmologies, non-singular expanding and inflationary cosmologies including 
a de Sitter (exponential) stage with positive scalar curvature.  
Non-singular contracting and deflationary cosmologies were also constructed.\\ \\ 
We analyse these cosmologies in detail with respect to the behaviour of the 
scale factors, the scalar curvature and the string-coupling.\\ \\
The sign of the scalar curvature turns out to be changed by the quantum corrections in 
oscillating cosmologies and evolves with time in the non-oscillating cases.\\ 
\\
Similarities between the two and three dimensional cases suggest a general 
picture for higher dimensional coset cosmologies :\\
(i) Anisotropy seems to be a generic unavoidable feature,\\
(ii) cosmological singularities are generically avoided and \\
(iii) it is possible to construct non-singular cosmologies where some 
spatial dimensions are experiencing inflation while the others experience 
deflation.\\
De Sitter stage can be achieved asymptotically at early times or late 
times, but there is not a conformal coset model of this type describing a 
de Sitter background globally.
\section{Quantum Field Theory, String Temperature and the String Phase of 
De Sitter Spacetime}
The density of mass levels ${\bf \rho}$(m) and the critical temperature 
for strings in de Sitter space-time were found in ref. \cite{ms1}.\\ \\
Quantum Field Theory (QFT) and string theory in de Sitter space have been 
compared in refs. \cite{ms1} and \cite{ms3}.\\ \\
A 'Dual'-transform is introduced which relates classical to quantum string 
lengths, and more generally, QFT and string domains.\\ \\
Interestingly, the string temperature in De Sitter space turns out to be the 
Dual transform of the QFT-Hawking-Gibbons temperature.\\ \\
The quantum back reaction problem for strings in de Sitter space is addressed 
selfconsistently in the framework of the 'string analogue' model (or 
thermodynamical approach), which is well suited to combine QFT and string 
studies \cite{ms1}, \cite{ms2}, \cite{ms3}.\\ \\
We find de Sitter space-time is a self-consistent solution of the semiclassical Einstein equations in this framework. Two branches for the scalar curvature 
R$_{(\pm)}$ show up : a classical, low curvature solution (-), and a quantum 
high curvature solution (+), entirely sustained by the strings. \\ \\
There is a maximal value for the curvature R$_{\mathrm {max}}$ due to the 
string back reaction.\\ \\
Interestingly, our Dual relation manifests itself in the back reaction 
solutions : the (-) branch is a classical phase for the geometry with intrinsic temperature given by the QFT-Hawking-Gibbons temperature.\\ \\
The (+) is a stringy phase for the geometry with temperature given by the 
intrinsic string de Sitter temperature.\\ \\
2 + 1 dimensions are considered, but conclusions hold generically in D 
dimensions.\\
\section{Hawking Radiation in String Theory and the String Phase of Black 
Holes}
The quantum string emission by Black Holes is computed in the framework of the 
'string analogue model' (or thermodynamical approach), which is well suited to 
combine QFT and string theory in curved backgrounds (particulary here, as black 
holes and strings posses intrinsic thermal features and temperatures).\\ \\
The QFT-Hawking temperature T$_{\mathrm{H}}$ is upper bounded by the string 
temperature T$_{\mathrm{S}}$ in the black hole background.\\ \\
The black hole emission spectrum is an incomplete gamma function of (
T$_{\mathrm{H}}$ - T$_{\mathrm{S}}$).\\ \\
For T$_{\mathrm{H}} \ll$ T$_{\mathrm{S}}$, the spectrum yields the QFT-Hawking emission.\\ \\
For T$_{\mathrm{H}}$ near to T$_{\mathrm{S}}$, it shows highly massive string 
states dominate the emission and undergo a typical string phase transition to 
a microscopic 'minimal' black hole of mass M$_{\mathrm{min}}$ or radius 
r$_{\mathrm{min}}$ (inversely proportional to T$_{\mathrm{S}}$) and string 
temperature T$_{\mathrm{S}}$.\\ \\
The semiclassical QFT black hole (of mass M and temperature T$_{\mathrm{H}}$) 
and the string black hole (of mass M$_{\mathrm{min}}$ and temperature T$_{\mathrm{S}}$) are mapped one into another by a 'Dual' transform which links 
semi classical-QFT and quantum string regimes.\\ \\
The string back reaction effect (selfconsistent black hole solution of the 
semiclassical Einstein equations with mass M$_{+}$ (radius r$_{+}$) and temperature T$_{+}$) is computed.\\ \\
Both, the QFT and string black hole regimes are well defined and bounded:  
r$_{\mathrm{min}} \le$ r$_{+} \le$ r$_{\mathrm{S}}$, 
M$_{\mathrm{min}} \le$ M$_{+} \le$ M$_{\mathrm{S}}$, 
T$_{\mathrm{min}} \le$ T$_{+} \le$ T$_{\mathrm{S}}$.\\ \\
The string 'minimal' black hole has a life time 
$\tau _{\mathrm{min}}$ = (K/Gh) T$_{\mathrm{S}}$ $^{-3}$. \\
 \section{New Dual Relation between Quantum Field Theory Regime and String 
Regime in Curved Backgrounds}
In ref.\cite{ms3} we introduced a R ``Dual'' transform which relates Quantum Field Theory and 
Quantum String regimes, both in a curved background.\\ \\
This operation maps the characteristic length of one regime into the other 
and, as a consequence, maps mass domains as well.\\ \\
The Hawking-Gibbons temperature and the string maximal or critical temperature 
are dual of each other.\\ \\
If back reaction of quantum matter is included, Quantum Field and Quantum 
String phases appear, and duality relations between them manifest as well.\\ \\
This Duality is shown in two relevant examples : Black Hole and de Sitter space times, and appears to be a generic feature, analogous to the ``wave-particle'' duality.\\
\section{New Coherent String States and Minimal Uncertainty Principle in 
String Theory.}
We study the properties of {\bf exact} (all level {\it k}) quantum coherent 
states in the context of string theory on a group manifold (WZWN models). \\ \\
Coherent states of WZWN models may help to solve the unitary problem : Having 
positive norm, they consistently describe the very massive string states 
(otherwise excluded by the spin-level condition).\\ \\
These states can be constructed by (at least) two alternative procedures : 
(i) as the exponential of the creation operator on the ground state, and (ii) 
as eigenstates of the annhilation operator. In the $ k \rightarrow \infty$ 
limit, all the known properties of ordinary coherent states of Quantum 
Mechanics are recovered. \\ \\
States (i) and (ii) (which are equivalent in the context of ordinary quantum 
mechanics and string theory in flat spacetime) are not equivalent in the 
context of WZWN models.\\ \\ The set (i) was constructed by Larsen and Sanchez 
in ref. \cite{ls}. The construction of states (ii) was provided in ref.  
\cite{ls2} by the same authors. We compare the two sets and discuss their 
properties. \\ \\
We analyze the uncertainty relation, and show that states (ii) satisfy 
automatically the {\it minimal uncertainty} condition for any {\it k}; 
they 
are thus {\it quasiclassical}, in some sense more classical than states 
(i) 
which 
only satisfy it in the $ k \rightarrow \infty$ limit.\\ The modification to the Heisenberg relation is given by $2 \mathcal{H}/k$, where $\mathcal{H}$ is 
connected to the string energy.\\

\end{document}